\documentstyle[12pt]{article}\textheight 230mm\textwidth 170mm

\begin{document}
\hspace*{11cm}IU-MSTP/15\\
\hspace*{11cm}hep-th/yy-mm-dd\\
\hspace*{11.6cm}August 1996
\begin{center}
 {\Large\bf Quantum Exchange Algebra and Locality \\
in Liouville Theory}
\end{center}

\vspace*{1cm}
\def\thefootnote{\fnsymbol{footnote}}
\begin{center}{\sc Takanori Fujiwara,}$^1$
{\sc Hiroshi Igarashi}
and  {\sc Yoshio Takimoto}
\end{center}
\vspace*{0.2cm}
\begin{center}
{\em $\ ^{1}$ Department of Physics, Ibaraki University,
Mito 310, Japan}\\
{\em Graduate School of Science and Engineering,
Ibaraki University, Mito 310, Japan}\\
\end{center}
\vfill
\begin{center}
{\large\sc Abstract}
\end{center}
\noindent
Exact operator solution for quantum Liouville theory is investigated
based on the canonical free field. Locality, the field equation 
and the canonical commutation relations are examined based on the 
exchange algebra hidden in the theory. The 
exact solution proposed by Otto and Weigt is shown to be correct 
to all order in the 
cosmological constant.



\newpage
\pagestyle{plain}

\noindent
The investigations of quantum Liouville theory [1-6]
have exposed 
rather rich canonical structures of the theory including 
the discovery of exact operator solutions [1,4-6]
and underlying 
quantum group symmetry [7-9]
as extensively 
studied by Gervais and his collaborators. Exact solutions not only 
enable us  detailed analyses of the system under considerations but 
also are useful in examining general frameworks which are 
considered to be applicable to the system.  
It is certainly desired to gain better understanding on the 
exact solutions. 

In this note we shall investigate the exact operator 
solution proposed by Otto and Weigt \cite{ow} and give a 
complete derivation 
of the formal expansion of the Liouville exponential in terms of 
a canonical free field by postulating (1) (pseudo-)conformal 
invariance, (2) locality and (3) the Liouville field 
equation. In ref. \cite{ow} the locality conditions were solved 
to third order in the cosmological constant and the general  
formula for the Liouville exponential was proposed. It turned 
out to be a kind of quantum deformation of the classical 
expansion \cite{Weigt}, suggesting underlying quantum group 
structure \cite{cgs}. We shall extend the result of ref. \cite{ow} 
to all order. Gervais and Schnittger \cite{gs-np-94} also gave a 
verification based on their algebraic scheme of chiral vertex 
operators \cite{cgs}. Here we basically follow the approach of 
refs. \cite{ow,kn} and show that there exists an interesting exchange 
algebra leading to the quantum deformation.\footnote{Exchange 
algebra in quantum Liouville theory has already been noted in 
refs. \cite{cgs,Babe-cmp-91}.} 

Let us start from the Liouville theory described by the action
\begin{eqnarray}
S=\frac{1}{\gamma^2}\int d\tau d\sigma\Biggl(\frac{1}{2}
\partial_\mu\varphi\partial^\mu\varphi
-\mu^2{\rm e}^\varphi\Biggr) ~,
\end{eqnarray}
where the coupling constant $\gamma^2$ is related to the Virasoro 
central charge 
by $\displaystyle{\frac{1}{\gamma^2}=\frac{25-D}{48\pi}}$. 
Using the parametrization of ref. \cite{dhj}, the classical 
solution to the Liouville field equation is given by
\begin{eqnarray}
 \varphi(\tau,\sigma)={\rm ln}
\frac{\partial_+A(x^+)\partial_-B(x^-)}{\Biggl(1+
\displaystyle{\frac{\mu^2}{8}}
A(x^+)B(x^-)\Biggr)^2} ~,\label{csol}
\end{eqnarray}
with $x^\pm=\tau\pm\sigma$.
We will restrict ourselves to the case where 
$A(x^+)$ and $B(x^-)$ can be chosen to satisfy the periodicity
$A(x^++2\pi)=\alpha A(x^+)$ and $B(x^--2\pi)=\alpha^{-1} 
B(x^-)$ with $\alpha\neq1$ \cite{gn,ow,kn}. They can be 
expressed in terms of a free field $\psi(\tau,\sigma)
=\psi^+(x^+)+\psi^-(x^-)$ as 
\begin{eqnarray}
  \label{cab}
  A(x^+)&=&C(P)\int_0^{2\pi}d\sigma'
{\rm e}^{\frac{1}{4}\gamma P
\epsilon(\sigma-\sigma')+\psi^+(\tau+\sigma')} ~,\nonumber\\ 
  B(x^-)&=&C(P)\int_0^{2\pi}d\sigma''
{\rm e}^{-\frac{1}{4}\gamma P\epsilon(\sigma-\sigma'')
+\psi^-(\tau-\sigma'')} ~,
\end{eqnarray}
where $\epsilon(\sigma)$ stands for stair-step function defined by
$\epsilon(\sigma)=2n+1$ for $2n\pi<\sigma<2(n+1)\pi$ with $n$ being an 
integer and $\displaystyle{C(P)=\Biggl(2{\rm sinh}
\frac{\gamma P}{4}\Biggr)^{-1}}$. We expand 
$\psi^\pm$ by 
\begin{eqnarray}
  \label{fexp}
  \psi^\pm(x^\pm)=\frac{\gamma}{2} Q+\frac{\gamma }{4\pi} Px^\pm
+\frac{i\gamma}{\sqrt{4\pi}}\sum_{n\neq 0}
\frac{1}{n}a^{(\pm)}_n{\rm e}^{-inx^\pm}~.
\end{eqnarray}
Then the zero mode momentum $P$ is related to 
$\alpha$ by $\ln \alpha=\frac{1}{2}\gamma P$. 

The key property of the theory is the existence of a canonical 
transformation between the interacting theory and the free 
field \cite{gn,dhj,bct,ow,kn}. Furthermore the improved stress tensor for the 
interacting theory can be 
expressed in terms of the free field as 
\begin{eqnarray}
  \label{stress}
  T_{\pm\pm}=\frac{2\pi}{\gamma^2}((\partial_\pm\psi)^2
-2\partial^2_\pm\psi) ~.
\end{eqnarray}
This satisfies the classical Virasoro algebra and generates pseudo-conformal 
transformations $x^\pm\rightarrow f^\pm(x^\pm)$, under which the Liouville 
exponential ${\rm e}^{\lambda\varphi}$ is transformed as a primary 
field of conformal weight $\lambda$. The conformality of the classical 
solution 
(\ref{csol}) can be more transparent if it is put in the form 
\begin{eqnarray}
  \label{csol2}
  {\rm e}^{\lambda\varphi}&=&{\rm e}^{\lambda\psi}\Biggl(1+\frac{\mu^2}{8}
S\Biggr)^{-2\lambda} \nonumber\\ 
&=&{\rm e}^{\lambda\psi}\sum_{m=0}^\infty \frac{(-1)^m}{m!}
\frac{\Gamma(2\lambda+m)}{\Gamma(2\lambda)}
\Biggl(\frac{\mu^2}{8}\Biggr)^mS^m~,
\end{eqnarray}
where we have introduced the ``screening charge'' 
$\displaystyle{S(\tau,\sigma)
=A(x^+)B(x^-)}$ as in ref. \cite{cgs}. The conformal weight of the 
Liouville exponential is carried by the free field exponential 
${\rm e}^{\lambda\psi}$, while $S$ is transformed as conformal 
field of weight zero. 

If there exists a quantum analogue of the 
canonical transformation from the interacting 
theory to the free field, we can 
define quantum theory via the free field for which the 
quantization is rather obvious. To establish this 
we must find an appropriate quantum formula for the Liouville 
field, and verify the canonical commutation relations and the 
field equation. 

In quantum theory we impose the commutation relations 
\begin{eqnarray}
  \label{ccr}
  [Q,P]=i~, \quad [a^{(+)}_n,a^{(+)}_m]=[a^{(-)}_n,a^{(-)}_m]
=n\delta_{n+m,0}~,
\end{eqnarray}
and introduce free field operator ordering for oscillatory modes 
and symmetric ordering for zero modes by $:{\rm e}^{aQ}f(P):\equiv
{\rm e}^{\frac{1}{2}aQ}f(P){\rm e}^{\frac{1}{2}aQ}$. We assume 
that the conformal symmetry remains intact and is generated by 
the normal-ordered stress tensor (\ref{stress}). 

Instead of dealing with $\varphi$ directly we will argue the Liouville 
exponential ${\rm e}^{\lambda\varphi}$ for arbitrary $\lambda$. 
Then the Liouville field can be defined by $\displaystyle{
\varphi\equiv\frac{d}{d\lambda}{\rm e}^{\lambda\varphi}
\Biggr\vert_{\lambda=0}}$. Since ${\rm e}^{\lambda\varphi}$ should 
have definite conformal weight, it can be expanded as 
\begin{eqnarray}
\label{qle}
  {\rm e}^{\lambda\varphi(\tau,\sigma)}=:{\rm e}^{\lambda\eta\psi
(\tau,\sigma)}:\sum_{m=0}^\infty \Biggl(\frac{\mu^2}{8}\Biggr)^m
C_m(\varpi,\lambda)(S(\tau,\sigma))^m ~,
\end{eqnarray}
where the parameter $\eta$ is chosen so that the lhs reduces to a 
(1,1) primary field for $\lambda=1$ and is given by 
$\displaystyle{\eta-\frac{\gamma^2\eta^2}{8\pi}=1}$ \cite{gn,bct,ow,kn}. 
It is convenient to define rescaled variables $\overline Q\equiv\gamma
\eta Q$ and $\displaystyle{\overline P\equiv\frac{1}{4}\gamma\eta P}$.
The screening 
operator $S$ must be of vanishing conformal weight and is assumed 
to be given by 
\begin{eqnarray}
\label{qsc}
S(\tau,\sigma)=\int_0^{2\pi}d\sigma'\int_0^{2\pi}d\sigma''
:{\rm e}^{\overline P\{\epsilon(\sigma-\sigma')
-\epsilon(\sigma-\sigma'')\}+\eta\psi^+(\tau+\sigma')
+\eta\psi^-(\tau-\sigma'')}: ~.
\end{eqnarray}
Since the stress tensor (\ref{stress}) commutes with $P$, 
the expansion coefficients $C_m(\varpi,\lambda)$ may depend 
on $P$ without conflicting the conformal symmetry. Here we 
use $\varpi$ defined by $\overline P=i\pi g\varpi$ with 
$\displaystyle{g\equiv\frac{\gamma^2\eta^2}{8\pi}}$ for later 
convenience and $C_0(\varpi,\lambda)=1$ is assumed. 
The effective Planck constant $h$ used in refs. \cite{cgs} 
corresponds to $\pi g$ in our case. The expansion (\ref{qle}) 
is formal in the sense that the operators $(S(\tau,\sigma))^m$ 
become ill-defined for fixed $g$  if $m$ is sufficiently large 
\cite{ow,cgs,kn}. 

The coefficients $C_m(\varpi,\lambda)$ can be determined by 
requiring locality  
\begin{eqnarray}
  \label{locality}
  [\;{\rm e}^{\lambda\varphi(0,\sigma)},
  {\rm e}^{\nu\varphi(0,\sigma')}\;]
  =0 ~. 
\end{eqnarray}
This is equivalent to the conditions 
\begin{eqnarray}
  \label{locality2}
  \sum_{m+m'=J}[V_\lambda(\sigma)C_m(\varpi,\lambda)(S(0,\sigma))^m,
  V_\nu(\sigma')C_{m'}(\varpi,\nu)(S(0,\sigma'))^{m'}]=0 ,
\end{eqnarray}
where we have introduced $V_\lambda(\sigma)
=:{\rm e}^{\lambda\eta\psi(0,\sigma)}:$ and 
the summation runs over nonnegative integers $m$, $m'$. 

To simplify (\ref{locality2}) we need to develop systematic way 
to expand the operators appearing in the lhs of (\ref{locality2})
by a set of independent operators. To find them we follow the idea 
of ref. \cite{cgs}.  We can restrict ourselves to the case 
$0<\sigma'<\sigma<2\pi$ without loss of generality and define
\begin{eqnarray}
  \label{ab}
  A_{\sigma\sigma'}&=&\int_0^{\sigma'} 
d\sigma''{\rm e}^{\frac{1}{2}\overline Q}
:{\rm e}^{\overline P+\frac{1}{\pi}\overline P
\sigma''+\eta\psi^+_{\rm osc}(\sigma'')}:
+\int_\sigma^{2\pi} 
d\sigma''{\rm e}^{\frac{1}{2}\overline Q}
:{\rm e}^{-\overline P+\frac{1}{\pi}\overline P
\sigma''+\eta\psi^+_{\rm osc}(\sigma'')}: \nonumber\\
  A_\Delta&=&\int_{\sigma'}^{\sigma} 
d\sigma''{\rm e}^{\frac{1}{2}\overline Q}
:{\rm e}^{\overline P+\frac{1}{\pi}\overline P
\sigma''+\eta\psi^+_{\rm osc}(\sigma'')}: \nonumber\\
  B_{\sigma\sigma'}&=&\int_0^{\sigma'} 
d\sigma''
:{\rm e}^{-\overline P-\frac{1}{\pi}\overline P
\sigma''+\eta\psi^-_{\rm osc}(-\sigma'')}:
{\rm e}^{\frac{1}{2}\overline Q}
+\int_\sigma^{2\pi} 
d\sigma''
:{\rm e}^{\overline P-\frac{1}{\pi}\overline P
\sigma''+\eta\psi^-_{\rm osc}(-\sigma'')}:
{\rm e}^{\frac{1}{2}\overline Q} \nonumber\\
  B_\Delta&=&\int_{\sigma'}^{\sigma} 
d\sigma''
:{\rm e}^{-\overline P-\frac{1}{\pi}\overline P
\sigma''+\eta\psi^-_{\rm osc}(-\sigma'')}:
{\rm e}^{\frac{1}{2}\overline Q}~,
\end{eqnarray}
where $\psi_{\rm osc}^\pm$ stand for the oscillator 
parts of $\psi^\pm$.  
Then the quantum analogue of (\ref{cab}) with omission of 
the overall coefficient are given by
\begin{eqnarray}
  \label{ab2}
  && A(\sigma)=A_{\sigma\sigma'}+A_\Delta~, \qquad
  B(-\sigma)=B_{\sigma\sigma'}+B_\Delta~, \nonumber\\
  && A(\sigma')=A_{\sigma\sigma'}+q^{-2\varpi-2}
  A_\Delta~, \qquad
  B(-\sigma')=B_{\sigma\sigma'}+q^{2\varpi}B_\Delta ~,
\end{eqnarray}
where we have introduced $q\equiv{\rm e}^{i\pi g}$. 
The screening operators appearing in (\ref{locality2}) 
are simply the products of these 
operators and, hence, can be 
expressed in terms of a set of four independent 
operators defined by 
\begin{eqnarray}
  \label{xyzw}
  x=A_{\sigma\sigma'}B_{\sigma\sigma'}~,\qquad 
  y=A_{\sigma\sigma'}B_\Delta~,\qquad
  z=A_\Delta B_{\sigma\sigma'}~,\qquad
  w=A_\Delta B_\Delta ~.
\end{eqnarray}
Though the operators (\ref{ab}) do not satisfy simple relations 
under interchanges in their products, (\ref{xyzw}) satisfies 
the exchange algebra 
\begin{eqnarray}
  \label{ea}
  yx=q^2xy ~, \quad zx=q^{-2}xz~, \quad wx=xw=q^{-2}yz=q^2zy,
  \quad wy=q^{-2}yw ~, \quad wz=q^2zw ~.
\end{eqnarray}
As for the exchange relations with the vertex operators, 
they commute with $V_\lambda(\sigma)$ and satisfy 
\begin{eqnarray}
 \label{eav}
xV_\lambda(\sigma')=V_\lambda(\sigma')x ~, ~~ 
yV_\lambda(\sigma')=q^{4\lambda}V_\lambda(\sigma')y ~, ~~
zV_\lambda(\sigma')=q^{-4\lambda}V_\lambda(\sigma')z ~, ~~
wV_\lambda(\sigma')=V_\lambda(\sigma')w ~. 
\end{eqnarray}
Using $[\overline Q,\varpi]=2$, we further find 
$x\varpi=(\varpi+2)x$, 
$V_\lambda(\sigma)\varpi=(\varpi+2\lambda)V_\lambda(\sigma)$ and similar 
relations for $y$, $z$, $w$ and $V_\lambda(\sigma')$. 

The exchange algebra (\ref{ea}) can be used to rearrange any product 
of operators (\ref{xyzw}) in a definite order. We can choose 
$x^ly^mw^n$ and $x^lz^mw^n$ $(l,m,n=0,1,2,\cdots)$ as a set of 
independent operators. There are just $(J+1)^2$ independent 
ones for $l+m+n=J$. 

We now come back to (\ref{locality2}). By interchanging the 
vertex operators and the screening charges, 
we can move the vertex operators to the left and then divide 
out them from the expression. We thus obtain  
\begin{eqnarray}
  \label{loc3}
  &&\sum_{m+m'=J}[ C_m(\varpi-2\nu,\lambda)C_{m'}(\varpi+2m,\nu)
  ({\cal S}(2\nu))^m({\cal S}(\varpi+1))^{m'} \nonumber\\
  &&\hskip 1.0cm -
C_{m}(\varpi-2\lambda,\nu)C_{m'}(\varpi+2m,\lambda)
  ({\cal S}(\varpi-2\lambda+1))^m({\cal S}(0))^{m'}]=0 ~,
\end{eqnarray}
where we have defined 
\begin{eqnarray}
  \label{cals}
  {\cal S}(\xi)=x+q^{2\xi}y+q^{-2\xi}z+w ~.
\end{eqnarray}
The lhs of (\ref{loc3}) is a homogeneous polynomial in the operators 
(\ref{xyzw}) of order $J$. Since there appear $(J+1)^2$ independent 
operators, it yields a set of redundant 
$(J+1)^2$  recurrence relations for $C_m(\varpi,\lambda)$. They are 
redundant in the sense that only the relations associated with the 
operators $x^{J-1}y$ and $y^J$ for $\lambda=\nu$ are sufficient 
to determine $C_m(\varpi,\lambda)$. Though this approach works well 
for small $J$ and reproduce the results of refs. \cite{ow,kn}, it is 
not so illuminating for arbitrary $J$. 

The condition (\ref{loc3}) can be simplified further if one notice 
that the exchange algebra (\ref{ea}) possesses a simple quantum mechanical 
realization \cite{cgs}. By using the same notation for the quantum mechanical 
realization, (\ref{ea}) is satisfied by the operators given by
\begin{eqnarray}
  \label{qmr}
  x=u{\rm e}^{\overline Q}~, \quad 
  y=u^2q^{-\varpi-2}{\rm e}^{\overline Q}~, \quad
  z=vq^{\varpi+2}{\rm e}^{\overline Q}~, \quad
  w=uv{\rm e}^{\overline Q}~,
\end{eqnarray}
where $u$ and $v$ are arbitrary parameters. In this realization the 
screening operator (\ref{cals}) can be written in a factorized form as 
\begin{eqnarray}
  \label{cals2}
  {\cal S}(\xi)=h(-\varpi+2\xi-2,\varpi-2\xi+2){\rm e}^{\overline Q} ~,
\end{eqnarray}
where $h(a,b)$ is given by 
\begin{eqnarray}
  \label{h}
  h(a,b)=(1+uq^a)(u+vq^b)~.
\end{eqnarray}
Putting (\ref{cals2}) into (\ref{loc3}) and moving all the 
${\rm e}^{\overline Q}$ to the right to eliminate them, we obtain 
\begin{eqnarray}
  \label{loc4}
  && \sum_{m+m'=J}\{C_m(\varpi-2\nu,\lambda)C_{m'}(\varpi+2m,\nu)
  \prod_{k=1}^m h(-\varpi+4\nu-2k,\varpi-4\nu+2k)
  \nonumber \\
  &&\hskip 3.5cm 
  \times\prod_{l=1}^{m'} h(\varpi+2m+2l-2,-\varpi-2m-2l+2) \nonumber\\
  && \hskip 1.2cm -C_m(\varpi+2m',\lambda)C_{m'}(\varpi-2\lambda,\nu)
  \prod_{k=1}^m h(-\varpi-2m'-2k,\varpi+2m'+2k)
  \nonumber \\
  &&\hskip 3.5cm
  \times\prod_{l=1}^{m'} h(\varpi-4\lambda+2l-2,-\varpi+4\lambda-2l+2)
  \}=0 ~.
\end{eqnarray}
The lhs is a polynomial in $u$, $v$ and contains only commuting 
variables. There arise $(J+1)^2$ independent terms proportional to 
$u^{J+k-l}v^l$ $(k,l=0,1,\cdots,J)$ in one-to-one correspondence 
with the $(J+1)^2$ independent operators appearing in (\ref{loc3}). 
Since (\ref{loc4}) holds identically for any $u$ and $v$, 
we can choose them arbitrarily so that the recurrence relations 
for $C_m(\varpi,\lambda)$ takes a simple form. The solution 
of (\ref{loc4}) proceeds in three steps. 

The first one is to resolve the $\varpi$ dependence. We 
choose $u=-q^{\varpi+2J}$ and $v=q^{4\nu+2J-2}$ for which  
$h(-\varpi-2J,\varpi+2J)=h(-\varpi+4\nu-2,\varpi-4\nu+2)=0$. 
In this case only the terms containing $C_J(\varpi,\nu)$ and 
$C_J(\varpi-2\lambda,\nu)$ remain nonvanishing
and (\ref{loc4}) reduces to 
\begin{eqnarray}
  \label{step1}
  \frac{C_J(\varpi,\nu)}{C_J(\varpi-2\lambda,\nu)}
    =\prod_{k=1}^J
    \frac{{\rm sinh}_q(\varpi-2\lambda+J+k-1)\;
      {\rm sinh}_q(\varpi-2\lambda-2\nu+k)}
    {{\rm sinh}_q(\varpi+J+k-1)\;{\rm sinh}_q(\varpi-2\nu+k)} ~,
\end{eqnarray}
where we have introduced $\displaystyle{{\rm sinh}_qx
\equiv\frac{q^x-q^{-x}}{2}}$. This immediately leads to 
\begin{eqnarray}
  \label{step2}
  C_m(\varpi,\lambda)=f_m(\lambda)\prod_{k=1}^m
  \frac{1}{{\rm sinh}_q(\varpi+m+k-1)\;{\rm sinh}_q(\varpi-2\lambda+k)} 
  \qquad (m=1,2,\cdots)~,
\end{eqnarray}
where $f_m(\lambda)$ are yet undetermined functions of $\lambda$. 

The next step is to find the $\lambda$ dependence of $f_m(\lambda)$.
For the purpose we consider 
the case $u=-q^{-\varpi+4\lambda}$ and $v=q^{-2\varpi+4\lambda+4\nu-2}$, 
where $h(\varpi-4\lambda,-\varpi+4\lambda)=h(-\varpi+4\nu-2,\varpi
-4\nu+2)=0$ and only the terms containing $C_J(\varpi,\lambda)$ and 
$C_J(\varpi,\nu)$ remain nonvanishing in (\ref{loc4}). This combined 
with  (\ref{step2}) gives the ratio $f_J(\varpi,\nu)$ with 
$f_J(\varpi,\lambda)$. We thus obtain 
\begin{eqnarray}
  \label{step3}
  f_m(\lambda)=c_m\prod_{k=1}^m{\rm sinh}_q(2\lambda+k-1) ~,
\end{eqnarray}
where $c_m$ is a $\lambda$ independent constant. 

The last step is to find $c_m$. Putting $u=-q^{\varpi+2J}$ and 
$v=q^{4\nu+2J-4}$, we get $h(-\varpi-2J,\varpi+2J)
=h(-\varpi+4\nu-4,\varpi-4\nu+4)=0$, 
for which (\ref{loc4})  
reduces to a recurrence relation between $c_J$ and $c_{J-1}$. This 
leads to 
\begin{eqnarray}
  \label{step4}
  c_m=\prod_{k=1}^m\frac{ia\;{\rm sinh}_q1}{{\rm sinh}_qk} ~,
\end{eqnarray}
where $c_1\equiv ia$ is a yet undetermined constant. 

Combining all these results, we finally obtain 
the expansion coefficients as 
\begin{eqnarray}
  \label{expc}
  C_m(\varpi,\lambda)=\prod_{k=1}^m\frac{ia\;{\rm sinh}_q1\;
    {\rm sinh}_q(2\lambda+k-1)}
  {{\rm sinh}_qk\;{\rm sinh}_q(\varpi+m+k-1)\;
    {\rm sinh}_q(\varpi-2\lambda+k)}
\end{eqnarray}
This can be shown to coincide with the results of ref. \cite{ow} 
for small $J$. We see that the requirement of conformal property 
and locality almost completely determine the expansion (\ref{qle}). 

Kazama and Nicolai \cite{kn} noted a solution to the locality 
conditions for $2\lambda g$, $2\nu g$ and $g$ being integers. In this 
case the screening charges and the vertex operators appearing in 
(\ref{locality2}) are mutually commuting as can be seen from (\ref{ea}) 
and (\ref{eav}). This implies that naive expansion of the classical 
solution holds true after the substitution $\psi\rightarrow\eta\psi$, 
provided the coefficients satisfy the periodicity 
$C_m(\varpi+g^{-1},\lambda)=C_m(\varpi,\lambda)$. This leads to the solution 
of ref. \cite{kn}, which cannot be obtained from (\ref{expc}) as a 
special case. 

We now turn to the field equation satisfied by $\varphi$. 
As mentioned previously, 
the Liouville field can be defined by differentiating 
(\ref{qle}) with respect to $\lambda$ and then putting 
$\lambda=0$. This leads to 
\begin{eqnarray}
  \label{lf}
  \varphi=\eta\psi+\sum_{m=1}^\infty\Biggl(\frac{\mu^2}{8}\Biggr)^m
  \frac{2\pi ig(ia\;{\rm sinh}_q1)^m}{{\rm sinh}_qm\;
    {\rm sinh}_q(\varpi+m)}
  \prod_{k=1}^{2m-1}\frac{1}{{\rm sinh}_q(\varpi+k)}S^m ~,
\end{eqnarray}
where use has been made of the relations 
\begin{eqnarray}
  \label{dlcm}
  C_m(\varpi,0)=\delta_{m,0} ~, \qquad 
  \frac{d}{d\lambda}C_m(\varpi,\lambda)\Biggr|_{\lambda=0}
  =\frac{2\pi ig(ia\;{\rm sinh}_q1)^m}{{\rm sinh}_qm\;
    {\rm sinh}_q(\varpi+m)}
  \prod_{k=1}^{2m-1}\frac{1}{{\rm sinh}_q(\varpi+k)} ~. 
\end{eqnarray}
Noting that the screening operator satisfies 
\begin{eqnarray}
  \label{dps}
  && \partial_\pm S^m=\frac{{\rm sinh}_qm\;{\rm sinh}_q(\varpi+m)}
  {{\rm sinh}_q1\;{\rm sinh}_q(\varpi+1)}\partial_\pm SS^{m-1} ~, 
  \nonumber \\
  && \partial_+\partial_-S^m=\Biggl(\frac{2\;{\rm sinh}_qm\;
  {\rm sinh}_q(\varpi+m)}{{\rm sinh}_q1}\Biggr)^2V_1S^{m-1} ~,
\end{eqnarray}
we obtain 
\begin{eqnarray}
  \label{ddphi}
  \partial_+\partial_-\varphi&=&-\pi ga\mu^2V_1\sum_{m=0}^\infty
  \Biggl(\frac{\mu^2}{8}\Biggr)^m(ia\;{\rm sinh}_q1)^m
  \frac{{\rm sinh}_q(m+1)\;{\rm sinh}_q(\varpi+m-1)}{{\rm sinh}_q1}
  \nonumber \\
  && \hskip 2.5cm 
  \times\prod_{k=1}^{2m+1}\frac{1}{{\rm sinh}_q(\varpi+k-2)}S^m ~.
\end{eqnarray}
Since the rhs just coincides to (\ref{qle}) multiplied by $-\pi ga\mu^2$ 
for $\lambda=1$, the Liouville equation is satisfied for the choice 
\begin{eqnarray}
  \label{a}
  4\pi ga=1
\end{eqnarray}
This determines the unknown constant $a$. 

What remains to show is the canonical commutation relations. 
This can be done along the strategy for the proof of 
locality. We first assume the expansion 
\begin{eqnarray}
  \label{phi}
  \varphi(\tau,\sigma)=\eta\psi(\tau,\sigma)
  +2\pi ig\sum_{m=1}^\infty\Biggl(\frac{\mu^2}{2}\Biggr)^m
  D_m(\varpi)(\overline S(\tau,\sigma))^m ~,
\end{eqnarray}
and then determine $D_m(\varpi)$ so as for the Liouville field 
to satisfy the commutation relations
\begin{eqnarray}
  \label{ccr2}
  [\varphi(0,\sigma),\partial_+\varphi(0,\sigma')]
  =[\eta\psi(0,\sigma),\eta\partial_+\psi(0,\sigma')] ~,
\end{eqnarray}
where we have introduced the notation $\overline S
=(2\;{\rm sinh}_q(\varpi+1))^{-2}S$.  
Using (\ref{dps}), we can equivalently write (\ref{ccr2}) as 
\begin{eqnarray}
  \label{ccr3}
  && \hskip -.5cm [\;\eta\psi(0,\sigma)\;,
  \frac{{\rm sinh}_qJ\;{\rm sinh}_q(\varpi+J)}{{\rm sinh}_q1\;
    {\rm sinh}_q(\varpi+1)}
  D_J(\varpi)\partial_+\overline S(0,\sigma')\overline S^{J-1}(0,\sigma')\;]
  +D_J(\varpi)[\;\overline S^J(0,\sigma)\;,\eta\partial_+\psi(0,\sigma')\;]
  \nonumber \\
  &&\hskip -.5cm +2\pi ig\sum_{m+m'=J}[\;D_m(\varpi)\overline S^m(0,\sigma)\;,
  \frac{{\rm sinh}_qm'\;{\rm sinh}_q(\varpi+m')}{{\rm sinh}_q1\;
    {\rm sinh}_q(\varpi+1)}
  D_{m'}(\varpi)\partial_+\overline S(0,\sigma')
  \overline S^{m'-1}(0,\sigma')\;]
  \nonumber \\
  &&=0 ~.
\end{eqnarray}
To simplify the lhs we need some commutation relations. For $0<\sigma'
<\sigma<2\pi$ we get
\begin{eqnarray}
  \label{psiS}
  && [\;\overline S(0,\sigma)\;,\eta\partial_+\psi(0,\sigma')]
  =-[\;\eta\psi(0,\sigma)\;,\partial_+\overline S(0,\sigma')]
  =\frac{2\pi ig}{{\rm sinh}_q(\varpi+1)}q^{\varpi+1}
  \partial_+\overline A(\sigma')\overline B(-\sigma)~,
  \nonumber \\
  && [\;\eta\psi(0,\sigma)\;,\overline S(0,\sigma')]
  =-\frac{2\pi ig}{{\rm sinh}_q(\varpi+1)}\{\;q^{\varpi+1}
  \overline A(\sigma')\overline B(-\sigma)+q^{-\varpi-1}
  \overline A(\sigma)\overline B(-\sigma')\} ~,
\end{eqnarray}
where we have introduced rescaled fields 
$\overline A=A(2\;{\rm sinh}_q\varpi)^{-1}$ and 
$\overline B=(2\;{\rm sinh}_q\varpi)^{-1}B$. 
These are the quantum analogue of (\ref{cab})
The extra factor 
$(2\;{\rm sinh}_q\varpi)^{-1}$ corresponds to the coefficient 
$C(P)$ of the classical formula (\ref{cab}). The commutation 
relations (\ref{psiS}) imply that the lhs of (\ref{ccr3}) is linear in 
$\partial_+A(\sigma')$, which can be moved to the left by using
the relation 
\begin{eqnarray}
  \label{sda}
  \overline S(0,\sigma)\partial_+A(\sigma')
  =\frac{{\rm sinh}_q(\varpi+3)}{{\rm sinh}_q(\varpi+1)}
  \partial_+A(\sigma')B(-\sigma)(A_{\sigma\sigma'}+q^{-4}A_\Delta) ~.
\end{eqnarray}
We then replace $\partial_+A(\sigma')$ with $A(\sigma')$ in the 
resulting formula. After these manipulations (\ref{ccr3}) can be 
converted into the form similar to (\ref{loc3}), to which we can 
apply the quantum mechanical realization (\ref{qmr}). Using the argument 
leading from (\ref{loc3}) to (\ref{loc4}), we obtain
\begin{eqnarray}
  \label{ccr5}
  && \Biggl\{i\pi g\frac{\partial}{\partial \varpi}\ln \Biggl(D_J(\varpi)
  \frac{{\rm sinh}_q(\varpi+J)}{{\rm sinh}_q(\varpi+1)}\Biggr)D_J(\varpi) 
  \nonumber \\ && 
  -D_J(\varpi)\sum_{k=0}^{J-1}\frac{q^{\varpi+2k+1}}
  {{\rm sinh}_q(\varpi+2k+2)}
  \frac{h(-\varpi-2k-2,-\varpi-2k)}{h(\varpi+2k,-\varpi-2k)}
  \nonumber \\ && 
  -D_J(\varpi)\sum_{k=1}^{J-1}\frac{q^{-\varpi-2k-1}}
  {{\rm sinh}_q(\varpi+2k+2)}
  \frac{h(\varpi+2k,\varpi+2k+2)}{h(\varpi+2k,-\varpi-2k)}\Biggr\}
  \nonumber \\ && \hskip .5cm
  \times\prod_{l=1}^Jh(\varpi+2l-2,-\varpi-2l+2)
  \nonumber \\ && 
  +\frac{D_J(\varpi)\;{\rm sinh}_q1}{{\rm sinh}_qJ\;{\rm sinh}_q(\varpi+J)}
  \sum_{m+m'=J}q^{\varpi+2m+1}h(-\varpi-2,-\varpi)
  \nonumber \\ &&\hskip .5cm
  \times\prod_{k=1}^{m-1}h(-\varpi-2k-2,\varpi-2k+2)
  \prod_{l=1}^{m'}h(-\varpi-2m-2l,\varpi+2m+2l)
  \nonumber \\ &&
  +\sum_{m+m'=J}\frac{{\rm sinh}_qm'\;{\rm sinh}_q(\varpi+J+m)}
  {{\rm sinh}_qJ\;{\rm sinh}_q(\varpi+J)}D_m(\varpi)D_{m'}(\varpi+2m)
  \nonumber \\ && \hskip .5cm
  \times h(-\varpi-2,-\varpi)h(\varpi+2m,\varpi+2m-2)
  \nonumber \\ && \hskip .5cm
  \times\prod_{k=1}^{m-1}h(-\varpi-2k-2,\varpi+2k-2)
  \prod_{l=1}^{m'-1}h(\varpi+2m+2l,-\varpi-2m-2l)
  \nonumber \\ &&
  -\sum_{m+m'=J}\frac{{\rm sinh}_qm\;{\rm sinh}_q(\varpi+m)}
  {{\rm sinh}_qJ\;{\rm sinh}_q(\varpi+J)}D_m(\varpi)D_{m'}(\varpi+2m)
  \nonumber \\ &&\hskip .5cm
  \times\prod_{k=1}^{m}h(\varpi+2k-2,-\varpi-2k+2)
  \prod_{l=1}^{m'}h(-\varpi-2m-2l,\varpi+2m+2l)=0 ~.
\end{eqnarray}
The recurrence relation for $D_m(\varpi)$ can be solved in the manner 
similar to (\ref{loc4}). We find 
\begin{eqnarray}
  \label{Dj}
  D_m(\varpi)=\frac{(ia\;{\rm sinh}_q1)^m{\rm sinh}_q(\varpi+2m)}
  {{\rm sinh}_qm\;{\rm sinh}_q(\varpi+m)}\prod_{k=1}^m
  \frac{{\rm sinh}_q(\varpi+2k-1)}{{\rm sinh}_q(\varpi+2k)} ~.
\end{eqnarray}
This coincides with the coefficients obtained from (\ref{lf}), 
establishing the commutation relation (\ref{ccr2}). The transformation 
(\ref{lf}) thus induces a operatorial canonical transformation. 

There remain some points to be clarified 
in our argument. For instance we should verify 
that the condition (\ref{loc4}) is satisfied for (\ref{expc}) 
for arbitrary $u$ and $v$. Related to this is the origin of the 
consistency of the redundant conditions for the expansion 
coefficients mentioned previousely. We will discuss these issues 
elsewhere. 

In conclusion it is possible to deform the classical Liouville 
field to satisfy the operator field equation and the canonical 
commutation relations by utilizing the degree of zero mode 
momentum. The formal expansion of the Liouville exponential 
given ref. \cite{ow} is correct to all order in the cosmological 
constant and for arbitrary parameters $\lambda$ and $g$.  

\eject

\vfill\end{document}